\documentclass[10pt,journal]{IEEEtranTCOM}
\usepackage[english]{babel}
\usepackage[usenames]{color}
\usepackage[cp1250]{inputenc}
\usepackage{amsfonts}
\usepackage{amsthm}
\usepackage{graphicx}
\usepackage{epsfig}
\usepackage{mathrsfs}
\usepackage{amsmath}
\usepackage{algorithm}
\usepackage{algorithmic}
\usepackage{hyperref}

\theoremstyle{plain}

\begin{document}
\newcommand{\bea}{\begin{eqnarray}}
\newcommand{\eea}{\end{eqnarray}}
\newcommand{\be}{\begin{equation}}
\newcommand{\ee}{\end{equation}}
\newcommand{\beas}{\begin{eqnarray*}}
\newcommand{\eeas}{\end{eqnarray*}}
\newcommand{\bs}{\backslash}
\newcommand{\bc}{\begin{center}}
\newcommand{\ec}{\end{center}}
\def\SC {\mathscr{C}}

\title{Exploiting context dependence for \\ image compression with upsampling}
\author{\IEEEauthorblockN{Jarek Duda}\\
\IEEEauthorblockA{Jagiellonian University,
Golebia 24, 31-007 Krakow, Poland,
Email: \emph{dudajar@gmail.com}}}
\maketitle
\begin{abstract}
Image compression with upsampling encodes information to succeedingly increase image resolution, for example by encoding differences in FUIF and JPEG XL. It is useful for progressive decoding, also often can improve compression ratio - both for lossless compression and e.g. DC coefficients of lossy. However, the currently used solutions rather do not exploit context dependence for encoding of such upscaling information. This article discusses simple inexpensive general techniques for this purpose, which allowed to save on average $0.645$ bits/difference (between $0.138$ and $1.489$) for the last upscaling for 48 standard $512\times 512$ grayscale 8bit images - compared to assumption of fixed Laplace distribution. Using least squares linear regression of context to predict center of Laplace distribution gave on average $0.393$ bits/difference savings. The remaining savings were obtained by additionally predicting width of this Laplace distribution, also using just the least squares linear regression.

For RGB images, optimization of color transform alone gave mean $\approx 4.6\%$ size reduction comparing to standard YCrCb if using fixed transform, $\approx 6.3\%$ if optimizing transform individually for each image. Then further mean $\approx 10\%$ reduction was obtained if predicting Laplace parameters based on context. The presented simple inexpensive general methodology can be also used for different types of data like DCT coefficients in lossy image compression.

\end{abstract}
\section{Introduction}
Beside lossy compression techniques like quantization, compression ratio depends on statistical modelling - predicting conditional probability distributions of values based on context, log-likelihoods of such models can be directly translated into savings in bits/value.

Laplace distribution (geometric when discretized) $\rho_{\mu b}(x)=\exp(-|x-\mu|/b)/2b$ turned out to be good universal  approximation for distribution of many types of values in data compression, like residues (errors from prediction), or AC coefficients of discrete cosine transforms (DCT). It has two parameters: center $\mu$ and width/scale parameter $b$. While context dependent prediction of value is often treated as estimator of $\mu$, the width parameter is often fixed. Rare example of predicting this width is LOCO-I/JPEG LS~\cite{loco}, which quantizes 3 dimensional context into 365 bins - not exploiting dependencies between them and rather being limited to low dimensional context.

\begin{figure}[t!]
    \centering
        \includegraphics{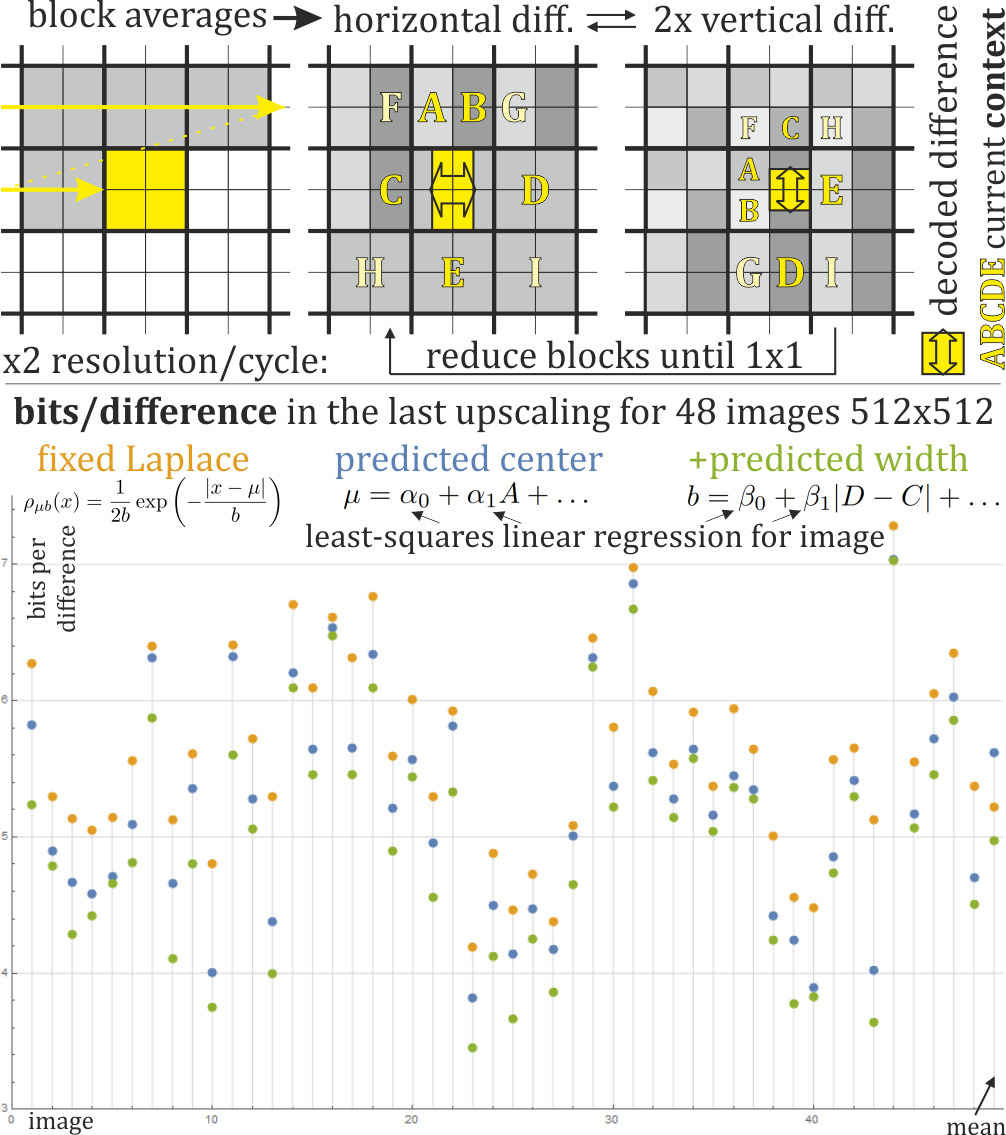}
        \caption{\textbf{Top}: some upsampling scheme - start with averages over some large square pixel blocks (or maybe the entire image), then scanning line by line uspample (increase resolution) alternately in horizontal/vertical direction (for RGB can be additionally split into 3 scans), using decoded sequence of differences. \textbf{Bottom}: evaluation on 48 grayscale 512x512 images (can be e.g. imagined as DC coefficients of 4096x4096 images) shown in Fig. \ref{img} of bits/difference for the last scan (twice more values than the previous one) for differences with complete context (without border). The highest orange dots assume Laplace distribution of fixed parameters. Lower blue dots show savings from predicting centers of Laplace distribution using linear combination of context values (fixed width optimized for all), with parameters chosen by least square linear regression individual for each image. The lowest green dots are for additionally also choosing width parameter based on context. Last column shown means over all 48 images.}
        \label{intr}
\end{figure}

We will focus on inexpensive general approach for predicting both centers and widths from \cite{param} - as just linear combinations of functions of context, with automatically optimized parameters e.g. with the least squares linear regression. It is computationally inexpensive, their parameters could be e.g. optimized for various region types, or even found by encoder for a given image and stored in the header. While it sounds natural for the centers, it might be  surprising that we can also predict width parameter this way: by MSE prediction of absolute values, what could be also alternatively done with more sophisticated models like neural networks.

This approach is applied here for image compression through upscaling: using sequence of differences to increase resolution. It is used for example in FUIF and JPEG XL~\cite{jxl} as "squeeze mode" of lossless image compression, however, they assume fixed $\mu=0$ Laplace distribution. As summarized in Fig. \ref{intr}, adding discussed inexpensive context-dependent prediction can bring essential savings: on average 0.645 bits/difference for the most costly: last scan. The previous scans have much lower number of values: twice per level. They got lower average saving: correspondingly 0.296, 0.225, 0.201 bits/difference for the previous three scans - these simple models are insufficient for higher level information, but can be helpful for filling details of textures - and this type of information often dominates bitstream.

Context dependence for symbol probability distribution is often exploited in the final symbol/bit sequence e.g. in CABAC~\cite{cabac} popular especially in video compression. However, such sequence looses spatial context information, which is crucial in image/video compression. Presented general approach can be also useful for exploiting context dependence for such situations, like modelling DCT coefficients using context e.g. of already decoded coefficients in current and neighboring blocks.

There were also analyzed RGB color images, for which optimization of color transform alone has turned out to bring a few percent improvements comparing to standard YCrCb.

\section{Methodology}
This main Section first briefly discusses "squeeze" upsampling approach, then approaches to predict center and width - their deeper discussion can be found in \cite{param}.
 \begin{figure}[t!]
    \centering
        \includegraphics{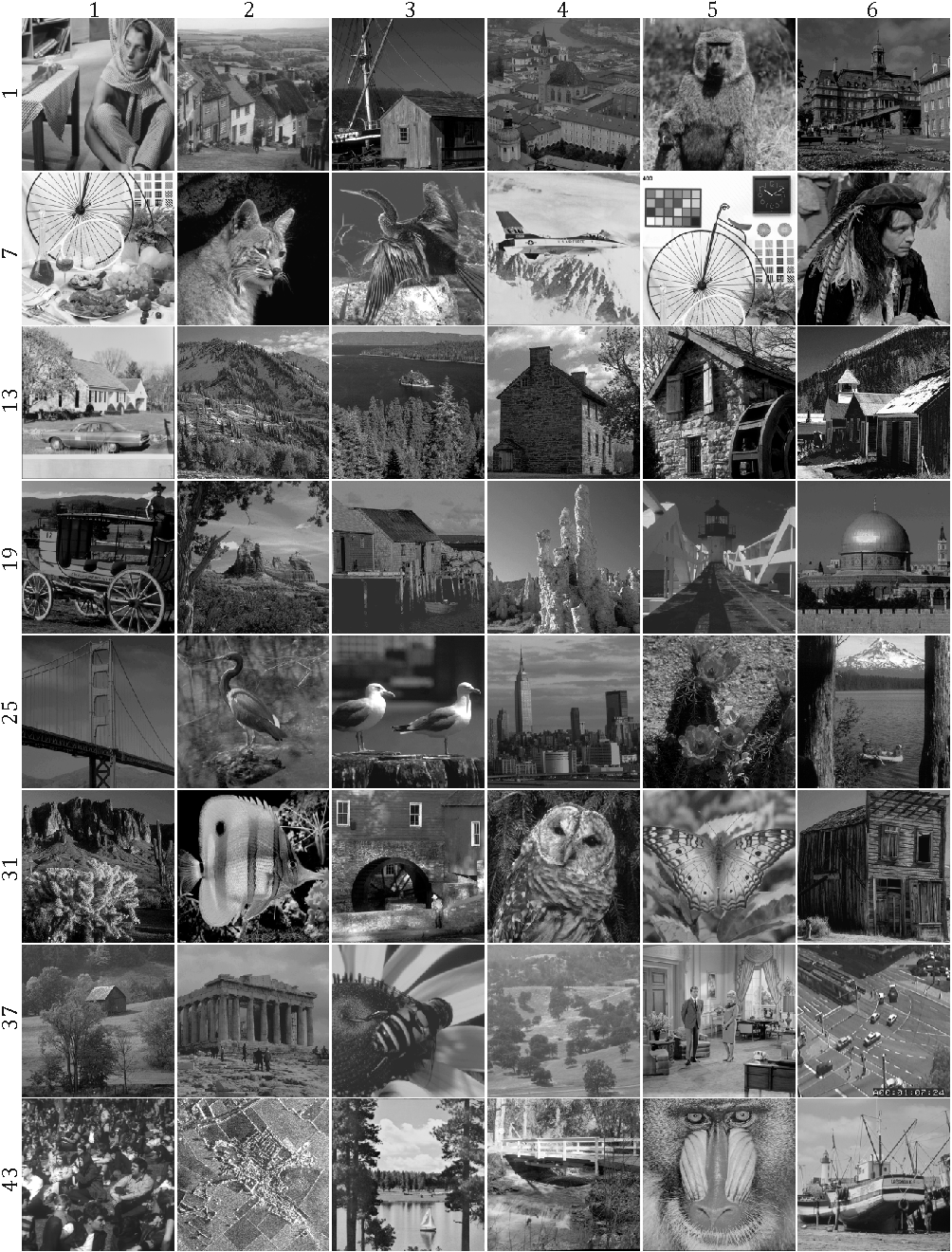}
        \caption{Dataset of 48 grayscale 8bit 512x512 images used in tests. Source: \url{http://decsai.ugr.es/cvg/CG/base.htm} .}
        \label{img}
\end{figure}
\subsection{Upsampling through "squeeze"}
There was used simple upsampling scheme presented in top of Fig. \ref{intr}, which can be seen as inspired by Haar wavelets~\cite{haar}: first store separately average over some square pixel regions (or even entire image), then succeedingly provide information about differences of averages of two subregions preferably of the same size, down to single pixel regions.

As we operate on discrete e.g. 8bit values, it would be convenient to maintain such range of integer values during upscaling, what can be done e.g. using "Squeeze" approach from Jon Sneyers' FUIF image compressor \footnote{https://github.com/cloudinary/fuif/blob/master/transform/squeeze.h}. Specifically, for $(u,v)$ higher resolution integer values, we use their $a$ average (integer but approximated) and $d$ difference:
\be a=\lfloor (u+v)/2 \rfloor\qquad\qquad d= u-v \ee
$(a,d)$ allow to uniquely determine $(u,v)$ as $\textrm{mod}(u+v,2)=\textrm{mod}(u-v,2)$, hence $u+v=2a+\textrm{mod}(d,2)$:
\be u= \lfloor (d +2a +\textrm{mod}(d,2))/2 \rfloor \qquad v=u-d \ee
We can for example scan line by line as in Fig. \ref{intr} and alternately upscale in horizontal and vertical direction, based on decoded sequence of differences $d$.

Statistics of these differences turn out agreeing well with Laplace distribution:

\be \rho_{\mu b}(x)=\frac{1}{2b}\exp\left(-\frac{|x-\mu|}{b} \right)\ee
The question is how to choose its parameters: center $\mu$ and width/scale parameter $b>0$? Standard approach is using fixed parameters. Their maximum likelihood estimation (MLE) for $(x_1,\ldots,x_n)$ sample is:
\be \mu=\textrm{median of }\{x_i\}\qquad b=\frac{1}{n}\sum_{i=1}^n |x_i-\mu|\ee

Let us discuss exploiting context dependence for better choice of parameters for a given position, what can lead to surprisingly large improvements as seen in Fig. \ref{intr} (also Fig. \ref{col}). We can use already decoded local context for this purpose, as in example in this figure, where yellow capital letters define values of context as averages over corresponding blocks.

\subsection{Predicting center $\mu$ from context $\textbf{c}=(A,B,\ldots)$}
While we could consider more sophisticated predictors including neural networks, considered basic family are linear predictors (of value from context $\textbf{c}=(A,B,\ldots)$):

\be \mu(\textbf{c})\equiv \mu=\alpha_0+\alpha_1 A +\alpha_2 B+\ldots\ee

A standard approach is finding fixed $(\alpha_j)$ parameters from interpolation: fit polynomial assuming some values in context positions, find its value in predicted position - getting a linear combination of context values.

A safer data-based approach is to directly optimize these parameters based on data: getting a single set of parameters optimized for a larger dataset, or better separate parameters for various region types (requiring e.g. a classifier). Parameters for tests here were optimized for a given image, for example to be found by encoder and stored in the header. The final solution should rather have some region classification with separate predictors, e.g. classified based only on context.

For  $(x_i)_{i=1..n}$ values and $(\textbf{c}_i)_{i=1..n}=(c_{ij})_{i=1..n, j=1..m}$ $m$-dimensional context (alternatively some functions on context), we can find parameters $(\alpha_j)$ minimizing
\be\arg\min_\alpha\  \sum_{i=1}^n  \left|x_i - \alpha_0- \sum_{j=1}^m \alpha_j c_{ij}\right|\qquad\  \left(\textrm{or }(x_i-\ldots)^2\right)\label{alpha}\ee
as MLE estimator of $\mu$ is median. From quantile regression~\cite{quantile} median can be predicted by minimizing mean $l^1$ norm - absolute value in (\ref{alpha}). However, MSE optimization: using squared $l^2$ norm instead is computationally less expensive and gives comparable evaluation - as it would rather have to be calculated by encoder in such applications, MSE optimization is used in tests here.

From experiments, the most crucial in predicting $\mu$ was $C-D$ difference (as in Fig. \ref{intr}) suggesting local gradient, which should be maintained between these positions especially in smooth regions. Also $A-B$ directly suggests this gradient: it is worth to include them into context. Finally the entire $(A,\ldots,I)$ size $m=9$ context was used in tests as still inexpensive and generally providing the best evaluation. For multiple channels we can add the already decoded into context.
\subsection{Predicting width parameter $b$ from context}

We can now subtract predicted $\mu\equiv \mu(\textbf{c})$ from values - denote such sequence as $(y_i)_{i=1..n}=(x_i-\mu(\textbf{c}_i))_{i=1..n}$. For these differences from prediction (residues) we could choose fixed width $b$, e.g. MLE: as mean $|y_i|$ - after quantization used for the blue dots in Fig. \ref{intr} evaluation.

We can improve by also predicting $b$ from the current context - again we could use more sophisticated models like neural networks, for simplicity in tests there was used linear combination of functions of context:

\be b(\textbf{c})\equiv b= \beta_0 +\beta_1 |A-B|+\ldots \ee
While for $\mu$ it is natural to directly use values from context in linear combinations, here we would like to estimate noise levels, which should be related to local gradient sizes, e.g. absolute differences of neighboring positions, generally some functions $(f_j)_{j=1..M}$ of context vectors $(\textbf{c}_i)_{i=1..n}$.

To inexpensively optimize $(\beta_j)$ for a chosen set of functions, remind that MLE estimation of $b$ is mean $|x-\mu|=|y|$. Observing that mean of values is the position minimizing mean square distance from these values, leads to heuristic:

\be\beta=\arg\min_\beta \ \sum_{i=1}^n \left(|y_i| - \beta_0- \sum_{j=1}^M \beta_j f_j(\textbf{c}_{i})\right)^2\label{beta}\ee

Which was used to get improvement between blue and green dots in Fig. \ref{intr} for context along gradient in decoded direction (plus absolute values of already decoded channels in current positron for RGB):
$$(1,|A-B|,|F-A|,|B-G|,|H-E|, |E-I|)$$
 We need to be careful here to ensure $b>0$, e.g. by enforcing all $\beta_j\geq 0$. In tests it was obtained by removing context leading to some negative $\beta$ and recalculating until all positive.

From entropy coding perspective, there should be prepared AC/ANS encoding tables for some quantized set of widths $b$ - one of them is chosen by $b$ predictor, such encoding step is applied to $y=x-\mu$ shifted (and rounded) value.

\section{Handling colors - optimizing transform}
Practically used images usually have 3 channels (or 4 with alpha). We can use the previously discussed approach for predicting centers and widths of Laplace distributions separately for all 3 channels, also increasing 3 times size of contexts (plus already decoded channels of current position) - what leads to mean $\approx 10\%$ additional savings. There was assumed decoding 3 channels at once, but it is also worth testing to split them into 3 succeeding scans with growing context.
\subsection{Aligning channels with axes for decorrelation}
In this subsection we will focus on improving color transform to decorrelate channels for agreement with assumption of 3 nearly independent Laplace distributions. Such change of basis (e.g. rotation) is popular inexpensive basic transformation. Its optimization alone turns out to allow for a few percent size reduction - directly for lossless compression maintaining e.g. 8bit value resolution, next subsection discusses adding quantization for lossy.

\begin{figure}[t!]
    \centering
        \includegraphics{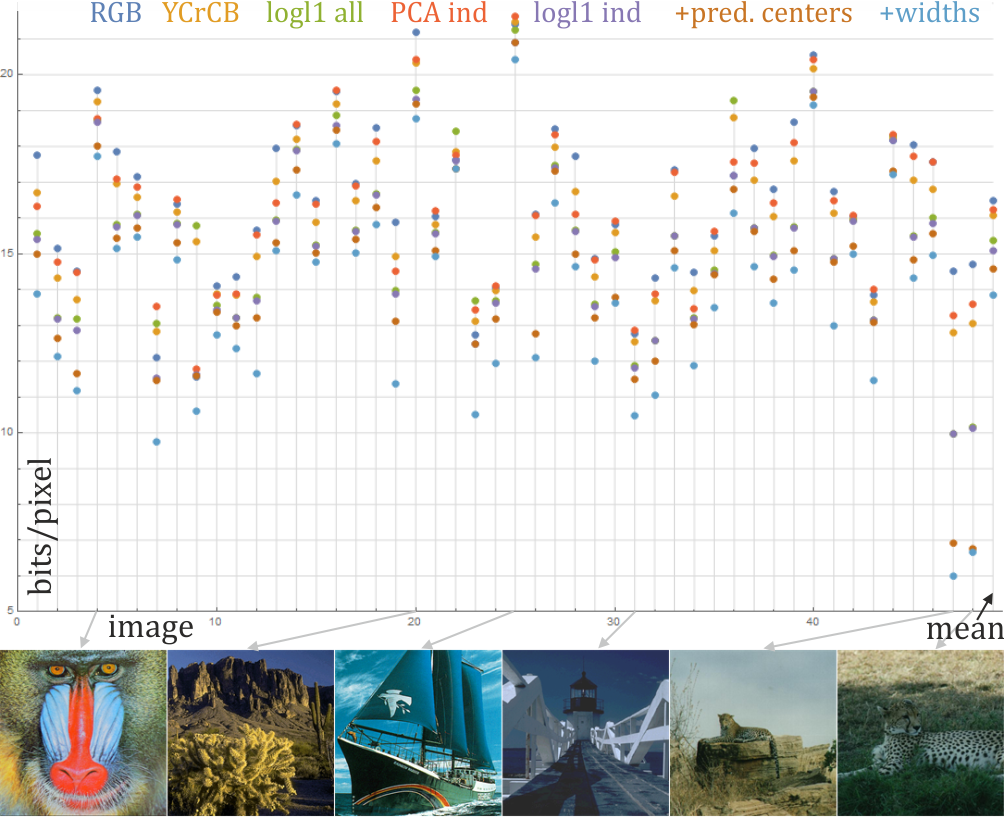}
        \caption{Evaluation: bit costs per pixel for 48 RGB 512x512 3x8bit images (source: \url{http://decsai.ugr.es/cvg/dbimagenes/c512.php} sorted alphabetically, can be imagined as DC coefficients of 4096x4096 images). After some rotation (RGB, YCrCb, log1 all, PCA ind, log1 ind), there were separately estimated 3 Laplace distributions for obtained channels - there are presented sums of their bits/pixel for 8bit values for lossless compression (originally 24 bits/pixel). RGB are for original 3 channels, YCrCb are for this standard transform. "PCA ind" use rotation found with PCA individually for each image. "logl1 ind/all" is for optimized discussed norm (\ref{logl1}): individually for each image, or using optimized common rotation for all 48 images. The remaining 3 improve on "logl1 ind": "+pred. centers" predicts value from neighbors and already decoded channels in current position, "+widths" additionally predicts widths of Laplace distributions from neighbors and absolute values of already decoded channels. Most-right column is mean for 48 images: correspondingly 16.4861, 16.0801, 15.369, 16.231, 15.098, 14.570, 13.859 bits/pixel.}
        \label{col}
\end{figure}

Let us now imagine that we have $X=\{x_{ij}\}_{i=1..n,j=1,2,3}$ values as $n\times 3$ matrix for $n$ RGB pixels. Evaluation presented in Fig. \ref{col} is for values coming from the last horizontal upscaling as discussed in the previous section, but such optimization can be also used for example for standard line scanning as in LOCO-I~\cite{loco} or \cite{param}. It might be beneficial to separately optimize it for various scenarios (including upscale direction, level).

Let us imagine (linear) color transform as $3\times 3$ matrix for RGB, in standard notation that each row describes linear combination for each transformed channel. While intuitively we would like an orthogonal matrix here, standardly used YCrCb transform\footnote{https://en.wikipedia.org/wiki/YCbCr} has matrix which is not orthogonal, leading to skewed value (e.g. 8bit) lattice. For fair comparison, let us focus on matrices having determinant equal 1 to maintain volume of cell of value/quantization lattice. Hence in evaluation there was used standard YCrCb matrix (ITU-R BT.601) multiplied by constant to get $\det=1$ (there was also tested orthonormalized, but it leads to similar evaluation):
$$O_{YCrCb}\approx 1.617479 \left(
  \begin{array}{ccc}
    0.299 & 0.587 & 0.114 \\
    -0.169 & -0.331 & 0.5 \\
    0.5 & -0.419 & -0.081 \\
  \end{array}
\right)$$
Looking natural approach is using PCA (principal component analysis) to choose the rotation (orthogonal), however, even individual optimization for each image has only lead to mediocre improvement. Intuitively, its $l^2$ optimization is proper for (multivariate) Gaussian distribution. In contrast, residues usually are from distribution close to Laplace, for which $l^1$ optimization is more appropriate, suggesting to use L1-PCA~\cite{L1PCA} instead.

\begin{figure}[t!]
    \centering
        \includegraphics{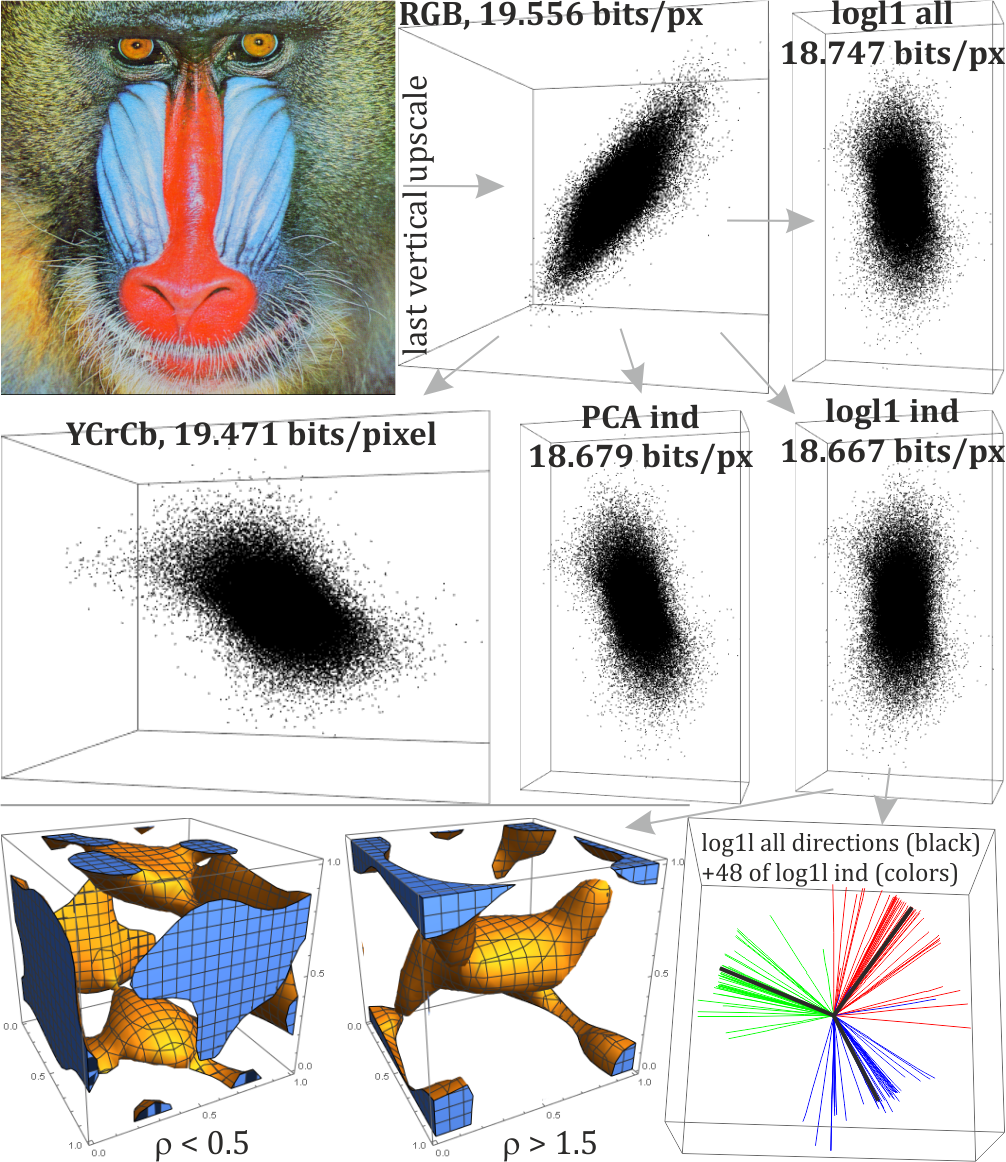}
        \caption{\textbf{Top}: example of set of values in discussed rotations for one of images, we can see improvement by decorrelation with aligning with axes - agreement of model as 3 nearly independent Laplace distributions. \textbf{Bottom}: some analysis of "logl1 ind" rotations - left: density below 0.5 and above 1.5 (modelled with polynomial after normalizing with Laplace distribution as in \cite{hcr}) showing type of dependencies - suggesting to increase width when already decoded channels have large absolute value. Right: obtained directions for 48 images.}
        \label{colsch}
\end{figure}

However, instead of guessing metrics, we can directly optimize for the actual problem. Entropy of width $b$ Laplace distribution is $\lg(2be)$, MLE estimator of $b$ is mean $|x-\mu|$. We can choose $\mu=0$, getting simplified evaluation to minimize (neglects $3\lg(2e/n)$ constant):
\be h(X)=\sum_j \lg\left(\sum_i |x_{ij}| \right)\qquad O_{logl1}=\underset{O:O^T O=\mathbf{1}}{\textrm{argmin}}\ h(XO^T) \label{logl1}\ee
Such $h(X)+3\lg(2e/n)$ is modelled differential entropy. There are better (but more expensive) evaluations: log-likelihood comparing dataset with assumed distribution, and finally entropy for quantized values - the last one is used in presented bits/pixel. Above simplified $e(X)$ evaluation could be generalized e.g. to use $|x_{ij}|^\kappa$ instead, especially for exponential power distribution~\cite{exp} family.

For (\ref{logl1}) orthogonal $3\times 3$ matrix optimization there was used quaternion representation $a^2+b^2+c^2+d^2=1$:
\begin{small}
$$\left(
  \begin{array}{ccc}
    a^2+b^2-c^2-d^2 & 2bc-2ad & 2bd+2ac \\
    2bc+2ad & a^2-b^2+c^2-d^2 & 2cd-2ab \\
    2bd-2ac & 2cd+2ab & a^2-b^2-c^2+d^2 \\
  \end{array}
\right)$$
\end{small}
There might be available some additional improvement if replacing orthogonality condition with weaker $\det=1$ as for YCrCb, to be tested in the future.

Such (\ref{logl1}) optimization for the last upscaling for combined datasets of all 48 images has lead to the following orthogonal color transform matrix:
$$O_{logl1}\approx\left(
  \begin{array}{ccc}
    0.515424 & 0.628419 & 0.582604 \\
    -0.806125 & 0.124939 & 0.578406 \\
    0.290691 & -0.767776 & 0.570980 \\
  \end{array}
\right)$$
It gave on average 1.117 bits/pixel lower cost than RGB, 0.711 bits/pixel lower cost than YCrCb, 0.862 bits/pixel than individually chosen PCA. Choosing rotation with logl1 individually for each image allowed for mean 0.217 bits/pixel additional reduction.

For evaluation in Fig. \ref{col}, after such individually chosen rotation using logl1, the lowest two dots correspond to further context-dependent prediction of centers of Laplace (brown, additional $\approx 0.5$ bits/pixel reduction), and then also widths (blue, additional $\approx 0.7$ bits/pixel reduction).

While exploiting this context dependence, it is worth to also include in the context already decoded channels. In prediction of the center, including previous channels contains some benefits of axis decorrelation - can  be used instead (a bit worse compression, not appropriate for lossy), has nearly negliglibe effect after decorrelation.

In width prediction, large absolute values of already decoded channels suggest to increase width of the following, what can be seen in density plots in the bottom of Fig. \ref{colsch}. Adding such absolute values into context for width prediction leads to $\approx 0.2-0.3$ bits/pixel additional savings.

\subsection{Lossy compression for decorrelated channels}
While the channel decorrelation savings were presented from perspective of lossless compression, it can be translated into lossy by adding quantization. There are promising vector quantization techniques like PVQ (pyramidal vector quantizer)~\cite{pvq}, however, currently used is mainly  uniform quantization we will focus on: choose some quantization coefficient $q$ (separately for various types of data e.g. channels), and encode $\textrm{round}(v/q)$ values instead of $v$, approximating the value with the closest node of step $q$ lattice $q\mathbb{Z}$.

In the previous subsection there was used differential entropy of Laplace distribution: $H=-\int_{-\infty}^{\infty} \rho(x) \lg(\rho(x)) dx = \lg(2be)$, which gets practical interpretation when approximated with Riemann integration for some quantization step $q$:
$$H=-\int_{-\infty}^{\infty} \rho(x) \lg(\rho(x)) dx \approx -\sum_{k\in\mathbb{Z}} q\rho(kq) \lg(\rho(kq))=$$
$$=-\sum_{k\in\mathbb{Z}} q\rho(kq) \lg(q\rho(kq)) + \sum_{k\in\mathbb{Z}} q\rho(kq) \lg(q)$$
The left hand side term is the standard entropy $h_q$ for probability distribution of quantization with step $q$. The right hand side term is approximately $\lg(q)$. So entropy for quantized continuous probability distribution is approximately the differential entropy plus $\lg(1/q)$:
\be h_q \approx H -\lg(q)\qquad \textrm{bits/value} \ee

In the previous subsection we have used $h(X)$, which after adding $3\lg(2e/n)$ constant can be seen as approximate evaluation of differential entropy for three independent Laplace distributions. Separately choosing quantization coefficients $q=\{q_1, q_2, q_3\}$ for the three channels would result in
\be \mathcal{H}(O,q)=h(XO^T) +3\lg(2e/n)-\lg(q_1 q_2 q_3)\quad \textrm{bits/pixel}\ee
approximate cost of storage for given transform $O$. Hence, if fixing the quantization coefficients, the optimized rotation would be as in the previous subsection.

Rotations optimized as previously (e.g. "logl1 all") usually have luminosity-like first vector (all positive coefficients), which should be stored with highest accuracy: low $q_1$. The remaining two channels correspond to determination of color (like Cr, Cb), can have higher $q_2,q_3$.

\subsection{Automatic optimization of qunatization coefficients}
It would be valuable to automatically optimize quantization coefficients, maybe together with rotation for higher level of optimization, based on some perceptually chosen distortion evaluation - especially for various compression settings, like optimizing compression ratio for a given distortion, or distortion for a given ratio.

A natural way to define such distortion in a chosen perceptual basis $P$ (e.g. $P=O_{YCrCb}$, not necessarily orthogonal or even square) is through mean-square error with corresponding positive  weights $d=(d_1,d_2,d_3)$ (e.g. larger for luminosity Y, lower for colors CrCb):

\be d(x) =  x^T P^T \left(
                               \begin{array}{ccc}
                                 d_1^2 & 0 & 0 \\
                                 0 & d_2^2 & 0 \\
                                 0 & 0 & d_3^2 \\
                               \end{array}
                             \right) P\, x =x^T D^2 x \ee
It allows to put both channels used for evaluation  and their weights into one symmetric $3\times 3$ matrix $D=P^T \textrm{diag}(d) P$ for $\textrm{diag}$ denoting diagonal matrix.

Now choosing quantization coefficients $q=(q_1,q_2,q_3)$ for transform matrix $O$, means errors proportional to $q_1$ in direction given by first row of $O$, $q_2$ in second, $q_3$ in third. So distortion evaluation for given $O,q$ is proportional to:
\be \mathcal{D}(O,q) = \sum_{j=1}^3 (q_j)^2 (O D^2 O^T)_{jj}=\left\| \textrm{diag}(q) O D \right\|^2_F\ \ee
for $\|M\|_F^2=\textrm{Tr}(MM^T)=\sum_{ij}M_{ij}^2$ Frobenius norm. 

We would like to optimize quantization $q$ to minimize distortion $\mathcal{D}$, what is easy to find for $P=O$ diagonal case
$$\underset{q:q_1 q_2 q_3=1}{\textrm{argmin}} \| \textrm{diag}(q) \textrm{diag}(d)\|_F = (d_1^{-1},d_2^{-1},d_3^{-1}) (d_1 d_2 d_3)^{1/3} $$
Intuitively, $q$ should be chosen to reduce differences in $d$, equalizing products for all three. Using different perceptual basis $P\neq O$, there appears additional orthogonal matrix between $\textrm{diag}(q)$ and $\textrm{diag}(d)$, what makes such equalization more difficult, for example leading to optimal $q_1=q_2=q_3$ if completely mixing all the coordinates. From perspective of this orthogonal matrix, distortion minimization gives it attraction to eigenbasis of $D$ matrix, strengthening with difference between $d$ weights.

We need to combine minimization of both $\mathcal{D}$ and 
$$ \mathcal{H}(O,q)=h(XO^T) +3\lg(2e/n)-\lg(q_1 q_2 q_3)\qquad \textrm{bits/pixel}$$
for what it is usually sufficient to optimize for $q_1 q_2 q_3=1$ and then rescale it using:
$$\mathcal{D}(O,\mu q)=\mu^2 \mathcal{D}(O,q)$$  \be\mathcal{H}(O,\mu q)=\mathcal{H}(O,q)-3\lg(\mu)\label{rescale}\ee
what allows for convenient rate-distortion control, e.g. increasing quantization $q$ twice, we save 3 bits/pixel, at cost of 4 times larger MSE distortion evaluation $\mathcal{D}$.\\

To summarise, it is more convenient to define perceptual evaluation in the same basis as rotation: $P=O$, e.g. previously found $O=O_{logl1}$. In this case optimal quantization is just $q_i=1/d_i$ times constant controlling rate-distortion tradeoff.

For general $P$ we can optimize both $O$ and $q$ simultaneously  by fixing value of $\mathcal{D}$ (instead of constraining $q_1 q_2 q_3$) and minimizing $\mathcal{H}$:
$$\underset{(O,q):\mathcal{D}(O,q)=\textrm{const}}{\textrm{argmin}} \mathcal{H}(O,q) $$
and then rescale $q$ with (\ref{rescale}) to control rate-distortion.

\section{Conclusions and further work}
There was presented application of general methodology from ~\cite{param} to data compression with upsampling - providing surprisingly large saving opportunity with low computational cost, which seems unexploited in current compressors.

This article only suggests basic tools, which can be improved e.g. with better choice of context, or functions of context especially for $b$ predictor. We can also use more sophisticated models like neural networks - preferably with $l^1$ optimization of distance from $x$ for $\mu$ predictor, and $l^2$ optimization of distance from $|x-\mu|$ for $b>0$ predictor (also ensuring positivity). However, such split of parameter prediction is an approximation, better compression ratios at larger computational cost could be obtained e.g. by further optimization of parameters directly maximizing log-likelihood for predicted conditional probability distributions.

There was also discussed optimization of color transform: while there is usually focus only on perceptual optimization, here it is combined with channel decorrelation, leading to additional a few percent savings. The remaining dependencies can be exploited in discussed width prediction - using that large absolute values in the previous channels should increase Laplace width in the succeeding.

Probably the most promising direction for further work is data-based automatic choice of separate predictors for various region types, e.g.  choosing one of models based only on the current context, or maybe mixing predictions from various models.

Another direction are other families of distributions, especially exponential power distribution~\cite{exp} (also containing Laplace distribution) - some initial tests provided $\approx 0.05$ bits/difference improvement.

We can also improve Laplace distribution model with further context dependent models of density as polynomial, like in \cite{hcr1}. Initial tests provided additional $\approx 0.15$ bits/difference improvements here, but these are relatively costly and large models, the question of their practicality here will require further investigation.

Order of decoding might bring some improvements due to different contexts of already decoded local situation, for example order of channels, or decoding 3 channels at once vs splitting into 3 succeeding scans - it is worth testing such various possibilities.

Finally, there should be also explored other applications of presented approaches, especially for DCT coefficients e.g. based on already decoded neighboring coefficients.

\bibliographystyle{IEEEtran}
\bibliography{cites}
\end{document}